\documentclass[journal=jacsat,manuscript=article]{achemso}
\usepackage{chemformula} 
\usepackage[T1]{fontenc} 
\usepackage{siunitx}
\DeclareSIUnit\angstrom{\text {Å}}
\usepackage{booktabs}
\usepackage{adjustbox}
\usepackage[colorlinks=true,linkcolor=black,anchorcolor=black,citecolor=black,filecolor=black,menucolor=black,runcolor=black,urlcolor=black]{hyperref}
\usepackage{placeins}
\usepackage{xcolor}
\usepackage{caption}
\usepackage{subcaption}
\usepackage{amssymb}
\usepackage[font=footnotesize,labelformat=simple]{subcaption}

\newcommand*{\citen}{}
\DeclareRobustCommand*{\citen}[1]{%
  \begingroup
    \romannumeral-`\x 
    \setcitestyle{numbers}%
    \cite{#1}%
  \endgroup
}
\renewcommand{\vec}[1]{\mathbf{#1}}
\newcommand{\spone}{\textsl{sp}}
\newcommand{\sptwo}{\textsl{sp}$^{2}$}
\newcommand{\spthree}{\textsl{sp}$^{3}$}

\author{Filip Vukovi\'c}
\email{vukovic@iap.tuwien.ac.at}
\affiliation[TU Wien]{Institute of Applied Physics, TU Wien, Vienna, Austria}
\altaffiliation{Contributed equally to this work}
\author{Anna Niggas}
\affiliation[TU Wien]{Institute of Applied Physics, TU Wien, Vienna, Austria}
\altaffiliation{Contributed equally to this work}
\author{Levin Mihlan}
\author{Zhen Yao}
\author{Armin Gölzhäuser}
\affiliation[Bielefeld University]{Faculty of Physics, Bielefeld University, Bielefeld, Germany}
\author{Louise Fréville}
\affiliation[Phelma INP Grenoble]{Phelma INP Grenbole, Grenoble, France}
\author{Vladislav Stroganov}
\author{Andrey Turchanin}
\affiliation[Jena University]{Institute of Physical Chemistry, Friedrich Schiller University Jena, Jena, Germany}
\author{J\"urgen Schnack}
\affiliation[Bielefeld University]{Faculty of Physics, Bielefeld University, Bielefeld, Germany}
\author{Nigel A. Marks}
\affiliation[Curtin University]
{Department of Physics, Curtin University, Perth, Australia} 
\author{Richard A. Wilhelm}
\affiliation[TU Wien]{Institute of Applied Physics, TU Wien, Vienna, Austria}

\title[]
{Revealing the innate sub-nanometer porous structure of carbon nanomembranes with molecular dynamics simulations and highly charged ion spectroscopy}
  
\begin{document}
%
%
%
%
%
%

\begin{abstract}
    Carbon nanomembranes (CNMs) are nanometer-thin disordered carbon materials
    that are suitable for a range of applications, from energy generation and storage, through to water filtration.
    The structure-property relationships of these nanomembranes are challenging to study using traditional experimental characterization techniques, primarily due to the radiation-sensitivity of the free-standing membrane.
    Highly-charged ion spectroscopy is a novel characterization method that is able to infer structural details of the carbon nanomembrane without concern of induced damage affecting the measurements.
    Here we employ molecular dynamics simulations to produce candidate structural models of terphenylthiol-based CNMs with varying degrees of nanoscale porosity, and compare predicted ion charge exchange data and tensile moduli to experiment.
    The results suggest that the in-vacuum CNM composition likely comprises a significant fraction of under-coordinated carbon, with an open sub-nanometer porous structure.
    Such a carbon network would be reactive in atmosphere and would be presumably stabilized by hydrogen and oxygen groups under atmospheric conditions.
\end{abstract}


\section{Introduction}

Nanometer thin carbon nanomembranes (CNMs) represent an intriguing class of thin-film materials with great potential in various nanotechnologies such as fuel cells\cite{Griffin2020}, energy storage\cite{Rajendran2021,Rajendran2023}
, photoactive electronics\cite{Tang2020,Kuellmer2022}, nanofiltration and water desalination\cite{YDB:AN18,vanDeursen2019,Yang2020,Stroganov2023,Stroganov2024}.
They are formed by irradiating self-assembled monolayers (SAMs) of organic molecules (e.g. biphenylthiol or terphenylthiol) grown on gold substrates with low energy electrons, resulting in a crosslinked soft film approximately one nanometer thick\cite{GSE:APL99,Turchanin2009,TBN:AM09,Turchanin2012,AVW:ASCN13,TuG:AM16,Tur:AdP17,DWN:PCCP19,WEG:2DM19}.
CNMs are generally considered to be non-crystalline soft carbon films\cite{YDB:AN18}, with a high selectivity for gas and water permeation\cite{Stroganov2023,Stroganov2024}.
Unlike graphite and related fullerenes, CNMs exhibit a lower tensile modulus between \SIrange{6}{12}{GPa} and ultimate tensile strengths between \SIrange{400}{700}{MPa}, depending on the precursor SAM\cite{Zhang2011,Zhang2021}.

In all applications, the structural characterization of the CNMs is crucial, as to a large extent the structure defines their functional properties. 
However, characterization of freestanding CNMs is significantly more challenging in comparison to typical 2D materials, such as graphene and MoS$_2$\cite{TuG:AM16,FeS:ACIE18}. 
For example, imaging CNMs via high-resolution transmission electron microscopy (HRTEM) induces graphitization due to the high electron beam energies, and hence hinders structural investigation\cite{Neumann2019}.
Scanning tunneling or atomic force microscopy (STM, AFM) can provide the structural information only for membranes on solid substrates.
Only recently, successful studies of CNMs at low temperatures were reported using non-contact AFM\cite{YDB:AN18}.
As such, there is a pressing need for characterization methods that can suitably probe the disordered structure of freestanding CNMs.

An alternative to these traditional experimental methods are keV beams of highly charged ions (HCIs), which have recently been introduced as a characterization technique to provide structural information of CNMs.\cite{Wil:JPCS20}
There are two key aspects to this method. 
Even though ions may modify the local site after impact, each ion carries with it information regarding the pristine sample state that it encountered, which is then interrogated after transmission at a detector.
The second is that total ion fluences that are used are extremely low, $<1000$\,ions/$\mu$m$^2$, therefore it is unlikely that ions impact regions that have already seen previous ion transmission.
HCIs are generated by removing several electrons from an atom.
Two key quantities define the HCI, the kinetic energy 
and the potential energy, which for the latter equals the sum of binding energies of all the removed electrons.
Upon approaching a surface, HCIs begin to neutralize via a two-center Auger-Meitner process known as interatomic Coulombic decay (ICD)~\cite{jahnke_interatomic_2015,jahnke_interatomic_2020,wilhelm_interatomic_2017}.
ICD is strongly distance dependent, going as $\sim 1/R^6$, with $R$ being the distance between the HCI and the material atoms~\cite{cederbaum_giant_1997,hemmerich_influence_2018}.
Hence, the charge exchange process is sensitive to the material thickness and sub-nanoscale structure. 

HCI transmission spectroscopy of CNMs can be used to study the structure of the membranes 
by analyzing outgoing charge states and scattering patterns.
High charge exchange (and large scattering angles) give insight regarding the thickness and density of the membranes.
Less charge exchange (and small scattering angles) give access to regions of lower density, and even atomic-scale voids, in the sample.
As with most spectroscopy methods, a model is required to infer sample properties from measured spectra.
In the case of HCI experiments, a model of the targets atomistic structure is necessary to link the experiment to HCI transmission simulations.


The disordered structure of CNMs poses a challenge not only for experimental characterization, but also theoretical descriptions as well, due in part to the limited experimental data.
Molecular dynamics (MD) simulations can be useful for elucidating structure-property relationships at the nanoscale, and have been successfully used to study carbon materials such as tetrahedral amorphous carbon\cite{Marks1996}, carbon foams\cite{Chen2021}, as well as fullerenes\cite{Powles2009}.
Recently, MD simulations have also been used to explore the SAM to CNM conversion process\cite{MrS:BN14,EGV:PRB21,SKY:BN22}. 
\citeauthor{EGV:PRB21} reported the first full MD simulations of carbon-only SAM to CNM conversion by imparting momentum impulses to the SAM and removing random molecules, considering biphenyl, terphenyl, and naphthalene thiol pre-cursor SAMs\cite{EGV:PRB21}.
They find that momentum transfer simulations are able to produce crosslinked CNM structures with some degree of porosity, however the predicted tensile moduli of the CNMs are significantly higher than experimental values.
While these prior efforts were able to produce promising CNM structures, they could not recover the experimental tensile modulus.

To address these limitations, we employ two distinct MD simulation approaches to produce candidate terphenylthiol precursor CNM structures.
The first method uses no spatial information of the precursor SAM. 
Instead, a pre-defined hole structure is enforced via modified particle dynamics, and used in conjunction with a liquid-quench and annealing simulation protocol to produce porous CNM structures with various degrees of porosity and carbon network development.
With this approach, CNM structures were created with varying degrees of sub-nano porosity and \textsl{sp}$^2$ fractions ranging from 40 \% to 90 \%.
The second approach is an extension of the momentum transfer simulation technique reported by \citeauthor{EGV:PRB21}, where discrete momenta are imparted to the carbon-only terphenylthiol SAM structure\cite{EGV:PRB21}.
We herein refer to the former as the `exclusion cylinder' method, and the later as `momentum transfer' simulations. 

The modelled CNM structures were then used as input for HCI transmission simulations.
These calculations were performed using the time-dependent potential (TDPot) method~\cite{Wilhelm2019}, which was modified to be compatible with non-2D target materials.
The modified TDPot description of HCI neutralization is based on the ICD model, and is able to predict angle-resolved exit charge state distributions in agreement with previous experiments of highly-charged Xe transmission through graphene\cite{Wilhelm2019}.
The simulated angle-resolved charge exchange spectra were then directly compared with experiment, thereby placing further constraints as to the true atomistic structure of the terphenyl-based CNMs.

\section{Methodology}
\label{sec-2}
\section{Experimental methods}

HCI transmission experiments were performed at the ion spectrometer at TU Wien, and are discussed in detail in Refs.~\citen{schwestka_versatile_2018} and \citen{niggas_coincidence_2022}.
A Dreebit Dresden EBIS-A~\cite{zschornack_compact_2008,kentsch_dresden_2002} was used, capable of producing Xe$^{q+}$ ions in charge states up to $q=44$ at a fixed acceleration potential of \SI{9}{kV}, resulting in ions with kinetic energies of $E_\textsubscript{kin}=9\times q$\,keV.
A Wien filter for charge-state selection, a set of deflection and focusing electro-optical components, and a pair of slits guide and shape the beam towards the terphenylthiol-based CNM samples.
The CNM samples used were prepared according to Refs.~\citen{eck2000generation}, and \citen{angelova2013universal}.

The HCI experiments were performed under ultra-high vacuum conditions ($<\SI{1E-9}{mbar}$) in transmission geometry, i.e. ions impinge normal to the surface and were detected after transmission on a 2D position-sensitive multichannel plate (MCP) detector. 
A slit and a pair of deflection plates with a field orthogonal to the slit were located between the sample and the MCP to separate distinct charge states for analysis. 
This arrangement permits study of scattering angle $\phi$ on the horizontal axis, and the charge state of ions along the vertical axis of the 2D MCP spectra.

\subsection{Contaminant removal by electronic excitation}


Standard surface cleaning techniques cannot be easily applied to very thin samples such as CNMs, as high-fluence sputtering can not only yield to contaminant removal but also to destruction of the sample itself.
Heating the sample is also only possible up to a limit, as the CNM undergoes a phase transition to graphene at temperatures above $\sim$\SI{900}{K}\cite{TBN:AM09,RWT:JPCC12,Turchanin2011}. 
To overcome these challenges, we apply slow HCI irradiation and spectroscopy at different incident Xe ion charge states.
With low charge states ($\lesssim q=10$) membranes were cleaned by means of soft desorption induced by electronic transitions\cite{tolk_desorption_1984,szymonski_new_1991}.
Higher charge states were then used to probe the angle-dependent charge exchange patterns and assess the cleanliness of the CNM\cite{niggas_role_2020}. 


HCI neutralization and the accompanying potential energy deposition introduces electronic excitations in the sample. 
It has been shown that small electronic excitations from low-energy electron irradiation and ultra-violet photons can lead to contaminant removal from single-layer graphene\cite{MaternaMikmekov2020,jia_toward_2016,tripathi_cleaning_2017}.
For HCIs, the potential energy desposited to the sample can be fine-tuned by selecting the incident charge state, with energies ranging from from several eV, up to hundreds of keV per ion.
We navigate a fine line between contaminant removal at low charge states and damaging of the sample with high charge states.
Depending on the electronic properties of the probed material, nanopore formation can occur upon HCI impact, i.e. pores with diameters up to several nanometers can form in the sample surface at the ion impact position\cite{RWS:APL13}.
Many materials, including CNMs, exhibit a potential energy threshold under which there is HCI-induced structural damage~\cite{wilhelm_threshold_2015,aumayr_single_2011,facsko_nanostructures_2009}.
We conclude that damage by Xe ions with $q=10$ and a kinetic energy of \SI{90}{keV} is sufficiently small as long as the applied fluence is less than $10^{12}$\,ions/cm$^2$, which corresponds to a fluence of $\lesssim 10^{-3}$ ions per carbon atom.
Note that conventional kinetic sputtering by momentum transfer from the ion to the carbon atoms is limited to only one or two atoms per incident ion, since an extended collision cascade that typically drives sputtering is not possible in a freestanding nanometer thin membrane. 

\begin{figure}
    \centering
    \includegraphics[]{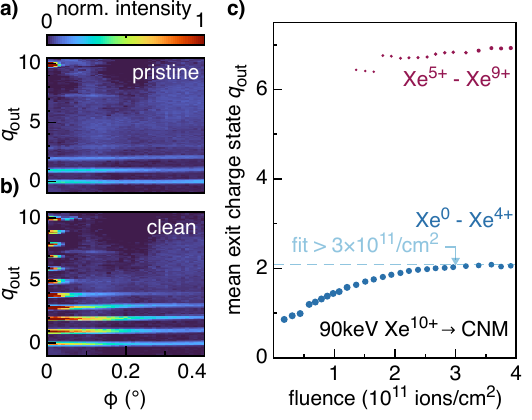}
    \caption{Exit charge state spectra of \SI{90}{keV} Xe$^{10+}$ after transmission through the CNM before (a) and after (b) contaminant removal through ion irradiation. (c) A fluence of $\sim 3\times 10^{11}$ ion impacts per cm\textsuperscript{2} is necessary to reach an equilibrium mean charge state of $q\textsubscript{out}\sim 2$.}
    \label{fig:cleaning}
\end{figure}

The exit charge state $q\textsubscript{out}$ distribution was monitored as a function of the applied fluence.
Figure~\ref{fig:cleaning}~(a) shows the exit charge state spectrum before cleaning, consisting mostly of neutrals and ions with $q\textsubscript{out}=1$ and 2, all scattered up to large angles.
This is a characteristic charge exchange pattern of thick materials where multiple scattering leads to large scattering angles and ions can fully neutralize~\cite{Creutzburg2021}.
Panel~(b) shows the exit charge state after sample preparation, where all charge states from incident $q=10$, through to neutrals are visible and the majority were scattered to scattering angles $\phi < 0.2^\circ$.
The appearance of more charge states and the overall decrease in scattering angles during sample cleaning are consistent with sample thinning, meaning contaminants were removed from the surface.

Figure~\ref{fig:cleaning}~(c) presents the mean charge state as a function of the applied fluence for the lower half and broader distribution (neutral Xe to Xe$^{4+}$ in blue), and upper half and smaller distribution (Xe$^{5+}$ to Xe$^{9+}$ in red).
Xe$^{10+}$ ion data was omitted from this analysis as it includes ions transmitted through micrometer-scale cracks\cite{niggas_role_2020}. 
For the low charge state distribution (blue in (c)), a continuous increase of the mean exit charge state with fluence can be observed, before it plateaus at $\sim 3\times10^{11}$\,ions/cm\textsuperscript{2} ($\sim 0.003$\,ions/nm$^2$).
These data support the idea that the membrane is not significantly damaged during ion irradiation, and that only physisorbed atomic and molecular species were removed.
If this were not the case, ions would constantly remove atoms from the material making it thinner and leading to a monotonic increase in the mean exit charge state.
Thus, after $\sim 3\times10^{11}$\,ions/cm\textsuperscript{2} the sample is deemed `\emph{clean}'.
Note that this fluence corresponds to \num{3.6E-5} ions per carbon atom in terphenyl-based CNMs.

Applying this cleaning procedure, all spectra exhibit a bimodal charge state distribution, which was also observed in previous studies that used nitro biphenyl thiol based CNMs~\cite{wilhelm_charge_2014,wilhelm_threshold_2015,wilhelm2016charge}.
Additional experimental data on nitro biphenyl thiol CNM cleaning as well as a comparison of nitro biphenyl thiol and terphenylthiol membranes (of different suppliers) can be found in the SI.
All results presented herein were achieved using CNMs cleaned according to the description above.

\section{Computational methods}
\subsection{Structural models of carbon nanomembranes via molecular dynamics simulations}

MD simulations were performed using a combination of LAMMPS \cite{LAMMPS}, and an in-house MD simulation code, both of which used the carbon EDIP potential to describe particle interactions\cite{Marks2000}.
The exclusion cylinder capability which is detailed in following section has been implemented in the stand-alone carbon EDIP MD code, and has been made available for use under the GPL 3.0 license.\cite{edip}
Structures were visualized using the OVITO software package \cite{ovito}, and analyses were performed using in-house codes.

\subsubsection{Exclusion cylinder molecular dynamics simulations}

The exclusion cylinder MD simulation method was developed as means of studying possible CNM structures by enforcing a distribution of voids during the structural evolution of the carbon network.
Specifically, we explored the effect of increasing areal hole density for a uniformly distributed set of holes with a fixed number of carbon atoms and unit cell dimensions (and therefore initial membrane thickness).
It should be emphasized that these simulations do not aim to model the SAM to CNM \emph{formation process}, and that we instead use a `means to an end' approach to produce candidate CNM structures with various pore arrangements.

For each exclusion cylinder configuration, $N_p$ cylinders were generated and placed about the $xy$ plane within the unit cell.
Positions of the circles were then moved using a simple hard-sphere like dynamics algorithm which terminated when no circles were within the specified inter-pore distance cutoff.
This cutoff was varied based on the cylinder count to maintain a roughly homogeneous distribution of exclusion cylinders.
These cylinders then defined the regions that atoms were not permitted to enter during certain stages of dynamics.
Nine unique configurations were used: 0, 10, 30, 50, 70, 90, 110, 130, and 150 cylinders, which correspond to an exclusion cylinder density of up to \SI{0.84}{nm^{-2}}.
The exclusion cylinders extend infinitely into the principle $z$ coordinate, i.e. perpendicular to the membrane surface, and obey the periodic boundary conditions in $x$ and $y$.
Periods of dynamics that incorporated the exclusion cylinders are referred to as \emph{region restricted} dynamics.
It should be noted that the exclusion cylinder algorithm treats the carbon atoms as point-particles with zero size, as such the diameter of the exclusion cylinders is not directly comparable to any experimental measure of porosity. 

Each simulation begins with the same number of carbon atoms (all other elements were omitted) and unit cell dimensions, i.e. \num{14040} atoms in a fully periodic (\numproduct{135 x 135 x 12}) \si{\angstrom^3} unit cell, which correspond to the experimental terphenylthiol SAM density of $\sim$\SI{1.3}{g/cc}.
Initial atom coordinates were randomly assigned within the allowed region for each configuration, so that no atoms were initially located within the exclusion cylinder locations.
In other words, each configuration of enforced exclusion cylinders begins from a unique starting configuration of randomly placed carbon atoms within the allowed volume. 
A mean exclusion cylinder radius of \SI{2.5}{\angstrom} was used, sampled from a gaussian distribution with $\sigma=0.5$.
This value was selected based on recent gas permeation experiments\cite{Stroganov2023,Naberezhnyi2019} that show with kinetic diameters under \SI{3.5}{\angstrom} are able to permeate through terphenyl-based CNMs, as well as prior ion selectivity experiments\cite{YHQ:AM20}, and water transport data
\cite{Stroganov2023,YDB:AN18,Dementyev2019,Dementyev2024}.
An integration timestep of \SI{0.35}{fs} was used for these simulations.

\begin{figure}
    \centering
    \includegraphics[width=1\linewidth]{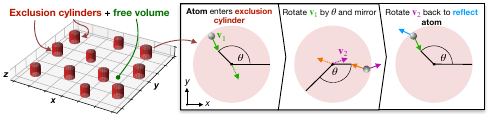}
    \caption{Schematic diagram of the specular reflection imposed on atoms that enter the exclusion cylinders during region restricted  dynamics.}
    \label{fig:spec}
\end{figure}

To prohibit atoms entering the exclusion regions during region restricted dynamics, the following routine was performed for all atoms and at each integration timestep.
After the forces and updated positions have been calculated, we check if $x,y$ coordinates of any atoms lie within the radius of any exclusion cylinders. If so, the velocity is reflected as
\begin{equation}
    \vec{v}' = 
    \begin{bmatrix}
         \cos \theta & -\sin \theta  \\
         \sin \theta & \cos \theta 
    \end{bmatrix}
    \begin{bmatrix}
        -1 & 0 \\
        0 & 1
    \end{bmatrix}
    \begin{bmatrix}
         \cos \theta & \sin \theta  \\
         -\sin \theta & \cos \theta 
    \end{bmatrix}
    \begin{bmatrix}
        v_x \\
        v_y
    \end{bmatrix},
\end{equation}
where $v_x$ and $v_y$ are the velocity components in $x$ and $y$, respectively, and $\theta$ is the angle between $v_x$ and the $x$ coordinate of the unit cell. 
Figure~\ref{fig:spec} schematically illustrates the specular reflection in several steps.
This specular reflection conserves total energy, and hence the standard equations of motion may be used.

The overall simulation protocol for each hole configuration was as follows.
Starting with the carbon atoms randomly packed in the unit cell, each hole configuration was subjected to region restricted dynamics with full 3D periodic boundary conditions for (i) \SI{3}{ps} in the constant particle, volume, and energy (\textsl{NVE}) ensemble, and (ii) an instant quench to a temperature of \SI{300}{K} which was held for \SI{3}{ps} using the Bussi thermostat\cite{Bussi2007}. 
Periodic boundary conditions in the $z$ direction were then switched off and a vacuum gap was used to create a thin slab with a surface normal to $z$.
This was followed by (iii) \SI{3}{ps} of simulation thermostatted to \SI{300}{K} to allow the surface to relax, (iv) a linear increase in thermostat temperature up to \SI{3000}{K} over \SI{10}{ps}, i.e. at a heating rate of \SI{270}{K/ps}, and then further maintained at \SI{3000}{K}.
Nine structural configurations were branched off after $0, 1, 4, 9, 16, 25, 36, 49,$ and \SI{64}{ps} of simulation time.
Once branched off, the structures are considered to be independent models.
For each of these simulation branches, the exclusion cylinder restrictions were then switched off and atoms were then permitted to move freely about the unit cell.
It is important to note that the annealing does not have any physical analogue to the experimental preparation of CNMs.
As all molecular simulations are limited by the time-scales that are accessible to them, higher temperatures are used to accelerate the evolution of the system so that many carbon atom configurations can be studied in a computationally tractable manner. 

Each branched frame was then immediately quenched to
\SI{300}{K}, and the pressure was maintained in the $x$ and $y$ unit cell directions at \SI{1}{atm} via the Nos\'e-Hoover barostat\cite{Parrinello1981} for a further \SI{300}{ps}.
This additional simulation time after quenching allows the structure to relieve excessive internal stresses before the tensile deformation simulations.
With the structure relaxed, the coordination fractions were calculated using a bond length cutoff of \SI{1.95}{\angstrom}.
The resultant bonding topology of the CNMs then consisted of a major single bond network and several disconnected small chains and isolated atoms.
All carbon atoms not connected to the dominant topology of the membrane were removed.
The cell dimensions in $x$ and $y$ remained within a few \AA ngstr\"oms of their initial values during this relaxation process.
A total of 81 unique candidate CNM structures were produced, spanning the high-temperature simulation time and exclusion cylinder count parameter space.

Tensile moduli for each CNM structure were predicted by deforming the $x$ or $y$ unit cell direction at an engineering strain rate of \SI{5E-3}{ps^{-1}}.
An integration timestep of \SI{0.1}{ps} was used for these tensile deformation simulations.
During the deformation, the unstrained cell dimension was allowed to vary to maintain \SI{1}{atm} of pressure. 
Stress-strain curves (Figures~S7 in the SI) were then used to calculate the tensile modulus by applying a linear fit to up to a strain of \num{0.02}, where the slope of the fit gives the tensile modulus\cite{Mihlan2025}.
Tensile moduli were calculated for both $x$ and $y$ deformation directions for each CNM structure, and hence the predicted modulus was taken as the mean of these two values.

\subsubsection{Momentum transfer simulations}

An alternative MD simulation approach was used to explore the formation of pores in terphenylthiol SAMs without explicit enforcement of holes.
The momentum transfer method is a refinement of the MD simulations reported by \citeauthor{EGV:PRB21}\cite{EGV:PRB21}, and begins with the assumption that the initial correlations in the SAM may partially carry through to the resultant CNM.
Specifically, these new simulations employ periodic boundary conditions in lateral ($x$ and $y$) membrane dimensions.
Details regarding the method can be found in Ref.~\citen{EGV:PRB21}, however, the process is briefly summarized below.

Each simulation begins with the same $\sim$\SI{135}{\angstrom} square unit cell.
Unlike the region restricted  simulations, initial coordinates of the \num{14040} atoms were arranged as a terphenyl SAM (with all elements other than carbon removed) on a gold substrate, oriented such that the SAM surface was normal to the $z$ coordinate.
Gold was modeled using a Lennard-Jones wall, using the C-Au parameters from Ref.~\citen{KDK:OBJ16}.
Figure~S8 in the SI presents an image of the initial SAM unit cell.

Irradiation of the SAM was modeled by a series of discrete momentum transfer events, where select groups of atoms were imparted with additional momentum to mimic collective electron impacts.
To this end, the SAM unit cell was divided into \num{1444} square regions, resulting in approximately \num{10} carbon atoms per region.
A single momentum transfer event imparts a primary downward force to all atoms within a randomly selected region, i.e. along the $-z$ direction. 
Regions that neighbor this primary region were subjected to a secondary planar force, that was directed away from the primary region.
For these simulations, a primary force of $\SI{50}{eV/\angstrom}$ was used.
Secondary force magnitudes and the number of force events were varied.
This applied force was exerted for a single timestep only, which was then followed by a relaxation period of regular \textsl{NVE} dynamics for a time of $t_r$.
For a given structure, a loop of $N_e$ impact events were performed, followed by a period of regular \textsl{NVE} dynamics for \SI{200}{fs}, and a further \SI{35}{ps} of thermostatted \textsl{NVT} dynamics at a temperature of \SI{300}{K}, using the Langevin thermostat\cite{Schneider1978}.
To mimic a cooling gradient through the gold surface, the thermostat is applied only in a defined region near the minimum of the Lennard-Jones potential.  Exemplary results of this method are shown in
Figure~S8 in the SI.

\subsection{Pore Detection}

A pore detection algorithm based on image processing techniques was developed to characterize the pore distribution in the produced CNM structures.
To quantify the pores that would be `visible' to incident ions, $xy$ coordinates of all atoms were binned into a spatial grid of pixels with a bin width of $\sim$\SI{0.25}{\angstrom} along the $x$ and $y$ axes.
Each pixel was then either marked as occupied or vacant, depending on if $xy$ coordinates of any atom was present within the pixel or not.
This array of pixels was then converted into a bitmap for further processing. 
At this stage, pores were visible as white areas and dense regions were shown as black areas.
Due to the relatively small bin widths, white noise was present in the dense regions.

Two morphological operations were used to remove noise pixels and determine the pore edges: `erosion' and `dilation'.
Erosion effectively swells occupied pixels such that the surrounding pixels are then set as occupied, thereby removing noise from the dense areas but also reduces pore areas.
A corrective dilation is then applied to the unoccupied pixels which is essentially the same operation as erosion, just applied to the unoccupied pixels.
Figure~S6 in the SI illustrates this process for a sample CNM structure.
With the bitmap treated, contours that define the pore edges were then determined.
The pixel count within closed contours was then used to calculate the pore area, and the number of closed contours yields the pore count.
Total CNM porosity was then taken as the percentage ratio of pore area to surface area of the unit cell.

\subsection{Highly-charged ion transmission simulations}

Charge exchange and scattering of HCIs transmitted through CNMs was modeled using the time-dependent potential TDPot approach of \citeauthor{Wilhelm2019}\cite{Wilhelm2019}.
The original TDPot method was developed to study stopping and scattering of highly charged ions transmitted through 2D materials, e.g. graphene, and as such had to be adapted for targets of finite thickness.
To this end, parameters that control the ICD rate \cite{Cederbaum1997,Wilhelm2017} were tuned by comparing the simulated ion charge state distributions to benchmark experimental data for single-layer, bi-layer, and tri-layer graphene (SLG, BLG, and TLG, respectively).
Figure~S2 in the SI presents the mean exit charge states comparison between simulation and the experimental fit functions for SLG, BLG, and TLG\cite{niggas_peeling_2021}.

Simulations of \SI{72}{keV} Xe$^{8+}$, \SI{135}{keV} Xe$^{15+}$, and \SI{180}{keV} Xe$^{20+}$ transmitted normal to the CNM surfaces were performed.
For each set of ion charge and kinetic energy parameters, \num{150000} individual ion trajectories starting at random $x$ and $y$ points within the unit cell were used to ensure sufficient sampling of the CNM.
Exit charge state of the ion and deflection angle data were used to create exit charge state histograms to compare with the experimental data.

\FloatBarrier
\section{Results and discussion}
\label{sec-3}

A total of 81 CNM structures were produced using the region restricted MD simulation approach.
Nine exclusion cylinder arrangements were used with a mean radius \SI{2.5}{\angstrom} and varying hole count from zero through to 150 cylinders (which corresponds to a hole density of 0.84 holes per \si{nm^2}).
As discussed in the Methods section, the term `\emph{hole}' refers to the enforced exclusion cylinders that were used as part of the simulation method, and so the terms `hole count', `hole density', and `hole diameters' refer only to this exclusion cylinder construction, and not the resultant membrane.

\begin{figure}[h]
    \centering
    \includegraphics[width=\textwidth]{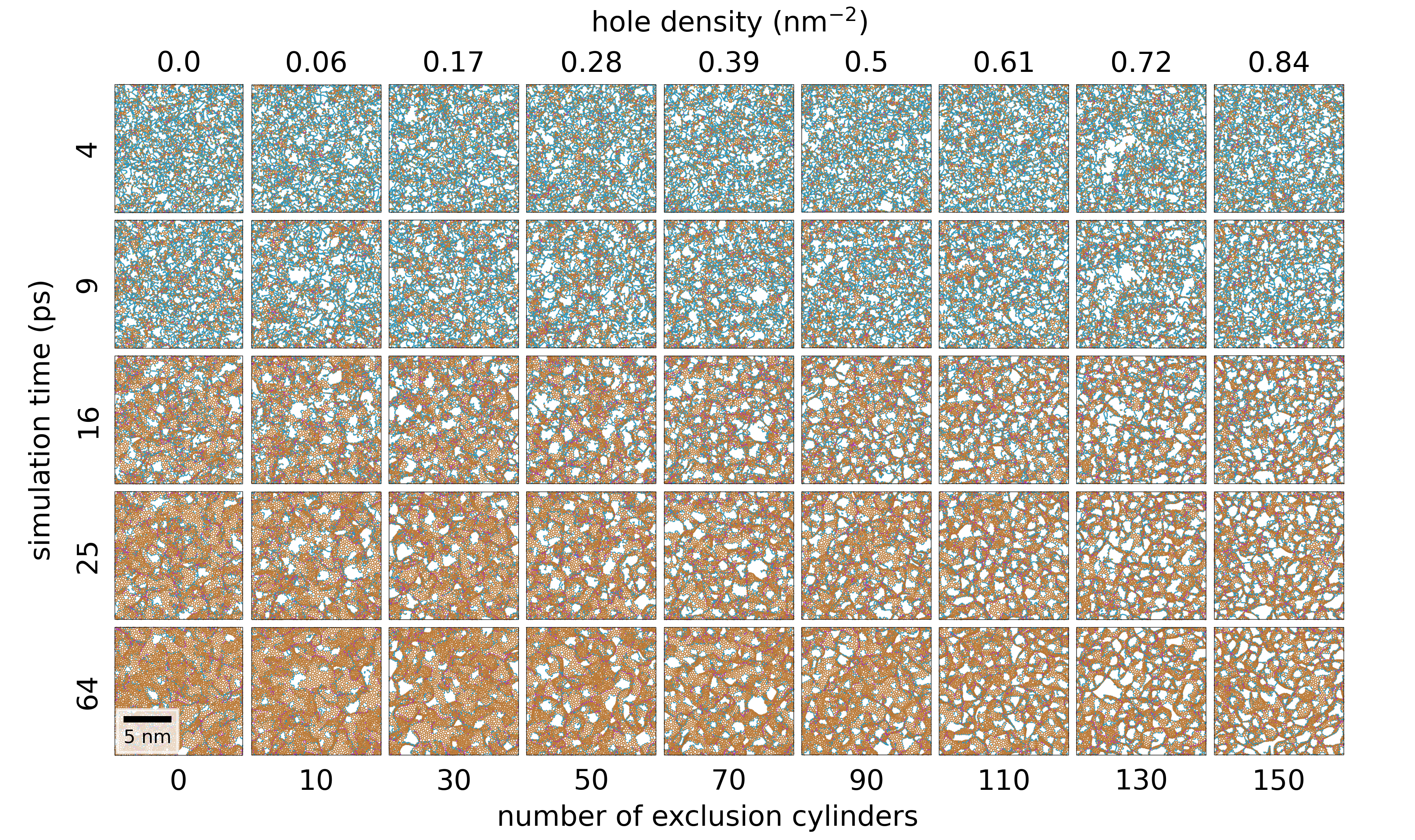}
    \caption{Planar top-down view of a selection of exclusion cylinder enforced carbon nanomembrane structures with atoms colored according to their coordination: \textsl{sp}$=$ blue, \textsl{sp}$^2=$ orange, \textsl{sp}$^3=$ purple, and single coordinated are shown in green}
    \label{fig:oz_cnm_matrix}
\end{figure}

Figure~\ref{fig:oz_cnm_matrix} presents a top-down view of all candidate CNM structures produced using the exclusion cylinder method, with atoms colored according to their coordination: \spone{} in blue, \sptwo{} in orange, and \spthree{} in purple.
All exclusion cylinder configurations began with roughly similar coordination fractions: \SI{50}{\%} \spone{}, \SI{40}{\%} \sptwo{}, with trace amounts of \spthree{}.
During simulation, the coordination fractions progress similarly.
There is a slight increase of \spone{} carbon up until \SI{9}{ps}, followed by a rapid decrease and plateauing out to below \SI{15}{\%}.
Note that here we use the term `\spone{}' imprecisely, as two-fold coordinated edge carbon atoms are in fact \sptwo{} radical in bonding character and only linear carbon chains would be true \spone{}.
However, for the sake of simplicity we do not make the distinction in our analysis.
For \sptwo{} carbon, there is a continuous increase at three distinct rates during simulation, and \spthree{} carbon increases slightly from $\sim$2 \%, up to 4 \%.
Graphs of the coordination fraction as a function of simulation time are shown in Figure~S3 in the SI.

\begin{figure}[!h]
    \centering
    \includegraphics[]{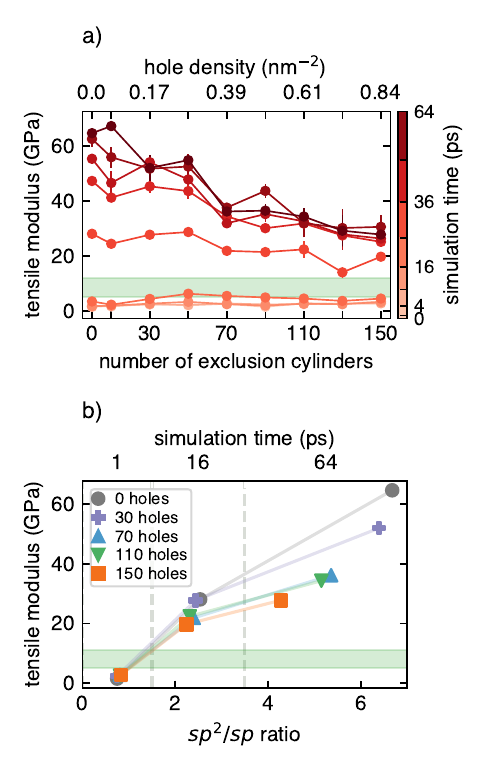}
    \caption{Predicted average tensile moduli of the CNM models as a function of the number of exclusion cylinders (panel a), where the color of each point represents the annealing time each structure has been subjected to. Each data point has been averaged over both in-plane directions, and error bars denote one standard deviation from the mean. The shaded green region indicates the reported experimental tensile modulus\cite{Zhang2011} range. Panel b) presents the same tensile data as a function of \textsl{sp}$^2$/\textsl{sp} for \SI{1}{ps}, \SI{16}{ps}, and \SI{64}{ps} simulated annealed structures, where the enforced cylinder counts are indicated by the different markers and colors.}
    \label{fig:tensile}
\end{figure}

Tensile moduli for all structures shown in Figure~\ref{fig:oz_cnm_matrix} were predicted from the linear regime of the CNM stress-strain response, where the tensile modulus is given by the slope of the linear fit.
Then CNM tensile modulus as a function of the number of exclusion cylinders is presented in Figure~\ref{fig:tensile}a), where the color of each line represents the simulation time of each of the structures.
The shaded green region denotes the typical experimental tensile modulus of terphenyl-based CNMs, which has been reported to be \SIrange{5}{12}{GPa}\cite{Zhang2011}.
Two overall trends are apparent, i) for each hole configuration the predicted modulus increases with simulation time, and ii) the spread of tensile modulus values due to annealing generally decreases for an increase in exclusion cylinder count.
Figure~\ref{fig:tensile}b) presents a sample of the tensile data for low, medium, and high annealing times (1, 16 and \SI{64}{ps}, respectively) as a function of the \textsl{sp}$^2$/\spone{} ratio.
Tensile modulus appears to be strongly correlated to the total \textsl{sp}$^2$ fraction of the membrane as the modulus generally increases with increasing \textsl{sp}$^2$ percentage, up to some threshold value for high hole densities.
Structures with low hole counts, up to about 31 holes, do not exhibit any plateau in the tensile modulus.

These data suggest that there are two ways of achieving a relatively (compared to that of pure graphite/graphene) low tensile modulus that is comparable to experimental CNM values.
One way to achieve this is to have a more disordered structure, comprising significant  amounts of under-coordinated carbon, $\gtrsim 30\%$ \spone{}.
However, carbon structures with high \textsl{sp}$^2$ content can also result in a low modulus given a sufficiently high hole density which then suppresses the in-plane stiffness of the membrane. 
Visual inspection of the highly annealed CNMs with 150 holes reveals that at high hole densities there is not enough space between pores for graphitic-like carbon to form lateral layers with the basal plane normal to the surface.
Instead, the carbon structure tends to form single- or few-layered fullerenes that wrap around the exclusion regions with the basal surfaces normal to the void.
In the extreme case, highly annealed membranes with high hole density resemble a forest of carbon nanotubes, inter-connected with rows of \textsl{sp}$^3$ carbon.
These results are consistent with previous theoretical reports of the transverse elastic constants for carbon nanotubes.
\citeauthor{Popov2000}\cite{Popov2000} calculated the elastic moduli of single-walled carbon nanotubes of various diameters using force-constant lattice dynamics, and report that nanotubes with a diameter of about \SI{10}{\angstrom} have a transverse tensile modulus of $\sim$\SI{10}{GPa}, which decreases for increasing tube diameters.

It is worth noting that a consequence of a membrane structure with large amounts of under-coordinated carbon is that it would be chemically reactive, and that dangling bonds would most likely be stabilized by hydrogen, and possibly water or other oxygen bearing groups when exposed to atmospheric conditions.
Our results therefore suggest that a non-reactive CNM with a tensile modulus of around \SI{10}{GPa} could only exist if the CNM resembled a crosslinked forest of nanotubes.
Such a structure is not compatible with the experimental Raman spectra data of terphenyl-based CNMs, which do not exhibit strong graphitic marker peaks, namely the D and G bands\cite{Zhang2018a}.

HCI transmission experiments were performed using terphenylthiol-based CNMs using \SI{72}{keV} Xe$^{8+}$, \SI{135}{keV} Xe$^{15+}$, and \SI{180}{keV} Xe$^{20+}$ incident ion beams.
TDPot simulations were then conducted for a range of the exclusion cylinder CNM structures shown in Figure~\ref{fig:oz_cnm_matrix}, yielding angle-resolved charge exchange spectra for incident Xe$^{q+}$ ions, which can be directly compared to experimental data.

\begin{figure}[!hb]
    \centering
    \includegraphics[width=0.9\linewidth]{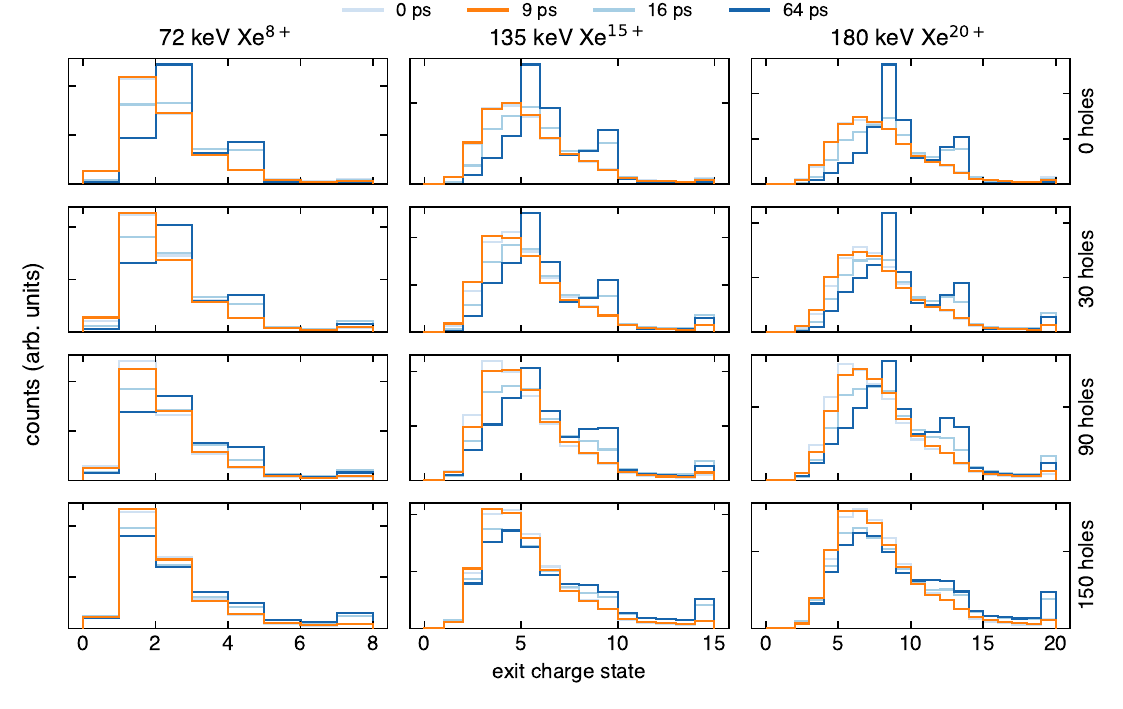}
    \caption{Simulated exit charge state for Xe ions incident on various hole-enforced nanomembrane structures. Structures that have been annealed for 0, 16, and \SI{64}{ps} are shown in darkening blue shades, and the \SI{9}{ps} structure is shown in orange.
    }
    \label{fig:simulated_cnm_charge_ex}
\end{figure}

Simulations of \SI{72}{keV} Xe$^{8+}$, \SI{135}{keV} Xe$^{15+}$, and \SI{180}{keV} Xe$^{20+}$ incident on the CNM model structures were performed.
Figure~\ref{fig:simulated_cnm_charge_ex} presents xenon exit charge state histograms for the zero, 30, 90, and 150 cylinder CNM structures for select simulation times.
Exit charge state distributions for structures annealed for 0, 16, and \SI{64}{ps} are shown in darkening shades of blue to illustrate the progressive effect of annealing.
Here, we highlight the structures annealed for \SI{9}{ps}, as after this time structures begin to exhibit a bi-modal exit charge state distribution. 
Lower values of exit charge (i.e. more neutral) are attributed to ions that pass through dense material and undergo significant neutralization, whereas the higher charge states are attributed to ions that pass through less-dense, or even open (relative to the ion ICD interaction radius), regions of the CNM.
The transition from a single skewed distribution to bi-modal distribution occurs when the model CNM structure develops significant layering, which then presents distinct single and bi-layer regions to incoming ions.
Additionally, as structures annealed for more than \SI{9}{ps} are in general much stiffer than the expected experimental tensile modulus.

\begin{figure}[!htb]
    \centering
    \begin{subfigure}[b]{0.9\textwidth}
        \centering
        \includegraphics[width=\textwidth]{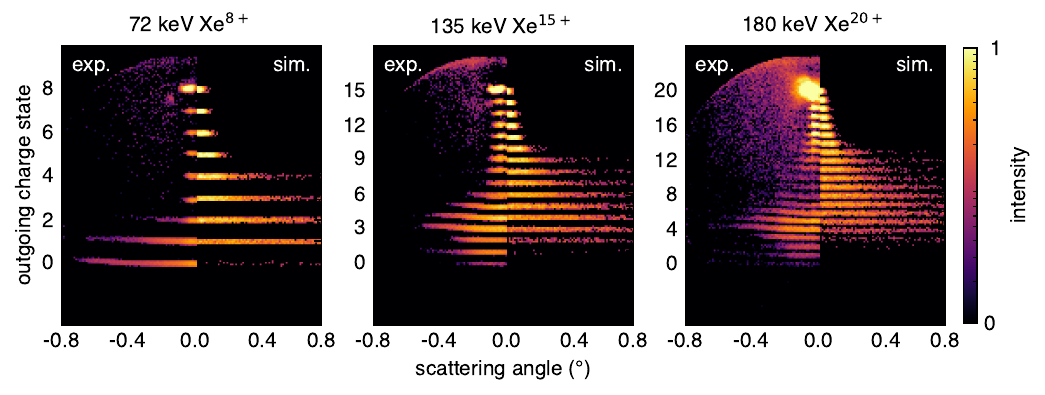}
        \caption{}
    \end{subfigure}
    \begin{subfigure}[b]{0.4\textwidth}
        \centering
        \includegraphics[width=\textwidth]{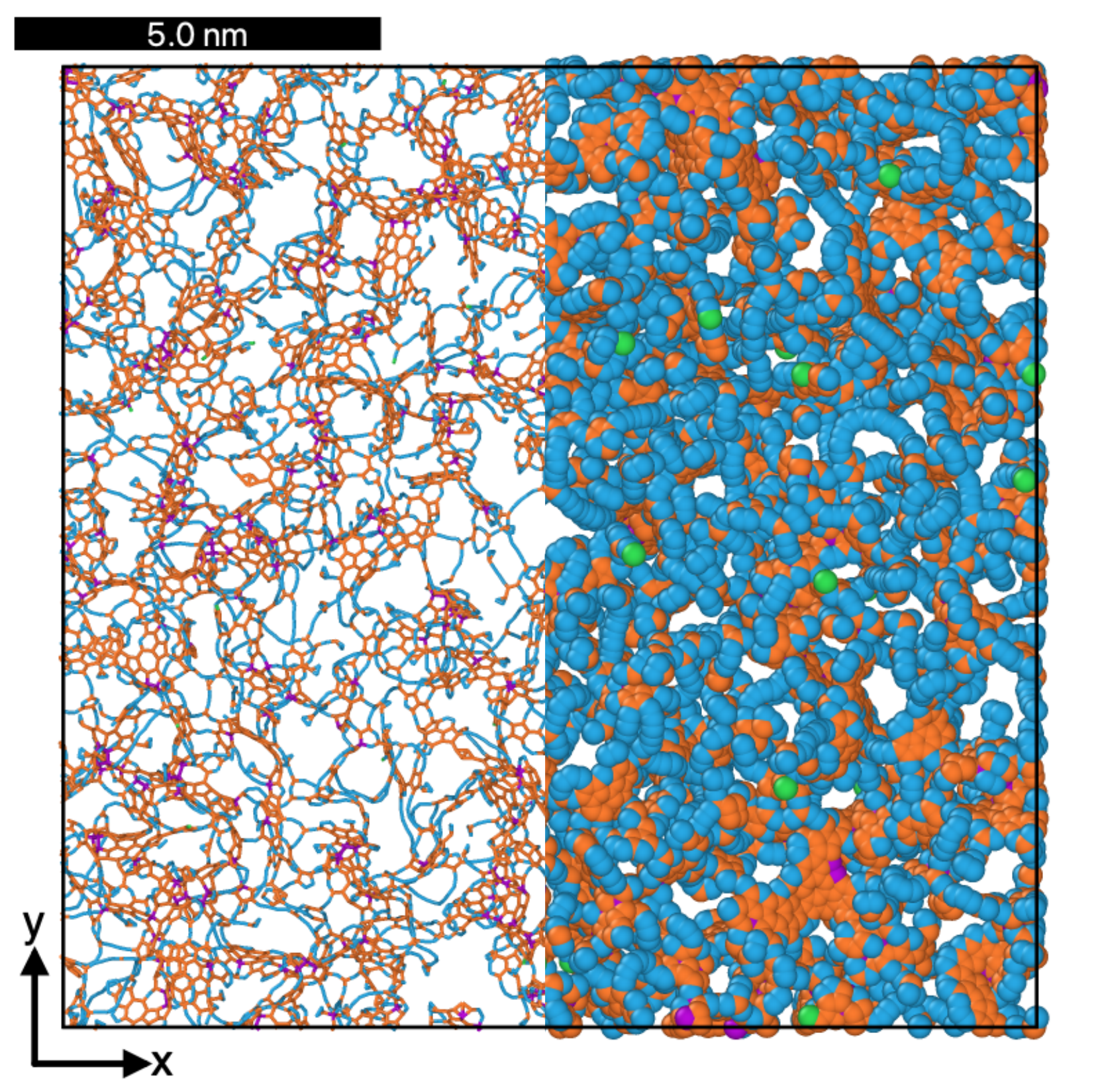}
        \caption{}
    \end{subfigure}
    \caption{Experimental (left half) and simulated (right half) charge exchange spectra through carbon nanomembranes (a) for \SI{72}{keV} Xe$^{8+}$, \SI{135}{keV} Xe$^{15+}$, and \SI{180}{keV} Xe$^{20+}$ from left to right. For the simulations, we used a 150 exclusion cylinder membrane annealed for \SI{9}{ps} shown in (b) in a top-down view, with atoms colored according to coordination as per Figure~\ref{fig:oz_cnm_matrix}. Here, the left half of the image is rendered using a bond-based visualization and the right half uses a van der Waals representation with an atom radius of \SI{1.7}{\angstrom}. Note that the scattering angles are different in experiment and simulation, details are given in the text.}
    \label{fig:final_charge_comp}
\end{figure}

The incident charge state is also prominent in many of the spectra shown in Figure~\ref{fig:simulated_cnm_charge_ex}.
This particular charge state increases with increasing exclusion cylinder count, which is due to the increase of open regions with the CNM.
These results show that the relative peak height of the incident charge state is sensitive to sub-nanoscale porosity, and that the ratio between this and mean of the lower charge state distribution could yield information regarding the total porosity of the CNM.
Unfortunately, it is currently not possible to extract such detail from experimental charge exchange spectra, as there is always an overwhelming contribution of incident charge states due to micro-scale cracks in the samples.
However, careful examination of high exit charge states may still yield valuable information regarding the porosity of the CNMs. 
In fact, Figure~\ref{fig:final_charge_comp} shows that the simulation captures the higher charge state distribution well, which is direct evidence of an innate sub-nanometer porosity.

Figure~\ref{fig:final_charge_comp}a) presents a comparison between the experimental charge exchange spectra, and the TDPot simulation result using the best fitting CNM structure which was the 150 exclusion cylinder CNM annealed for \SI{9}{ps}, and panel b) presents a top-down view of the CNM with the usual coordination based color scheme with two types of visual representation shown: a bond-based visualization on the left, and a van der Waals representation with a carbon radius of \SI{1.7}{\angstrom} on the right.
Although the simulations capture key aspects of the experimental charge state distribution, namely a narrow distribution scattering for higher charge states as well as larger scattering for lower charge states, some discrepancies are present.
In general the simulations exhibit less overall neutralization compared to experiment.
For Xe$^{20+}$ there is a more pronounced separation between the upper and lower charge state distributions in the experiment, compared to the simulations.

These data in combination with the predicted tensile modulus, suggest that low simulation time CNMs with higher numbers of enforced cylinders may be good candidate models for terphenylthiol-based CNM.
The low charge state distribution can be linked to ions transmitting through densified parts of the CNM.
The densification of the membrane and minimal carbon loss ($\lesssim$ 5\%) already implies a certain porosity of the resulting membrane and ion charge exchange in these parts of the membrane is therefore indirect evidence of sub-nanometer porosity.
Our TDPot simulations (cf. Figure~\ref{fig:simulated_cnm_charge_ex}) in comparison to angle-resolved experimental data (cf. Figure~\ref{fig:final_charge_comp}) allows a quantification which potential structure yields the best match also for the densified part of the membrane.

While the exclusion cylinder enforced MD simulations were able to provide insights regarding the possible atomistic structure of CNMs, the question remains whether it is possible to form a porous carbon network without the constraint of forced cylinder geometry.
To explore this idea, additional MD simulations were performed using the momentum transfer method, where localized force events and light annealing are used to induce pore formation in a carbon-only terphenyl SAM.
Here, two simulation parameters were varied to systematically study the formation of pores: i) the so-called secondary force applied to the SAM at each impact event (from \SI{300}{eV/\angstrom} to \SI{550}{eV/\angstrom}), and ii) the number of impact events (30 through to 50). 
The former will be referred to as `force-varied' and the latter as `event-varied'.
The resultant pore structure of these structures was then analyzed, as well as for the hole-enforced structures.

\begin{figure}[!h]
    \centering
    \includegraphics[width=0.5\textwidth]{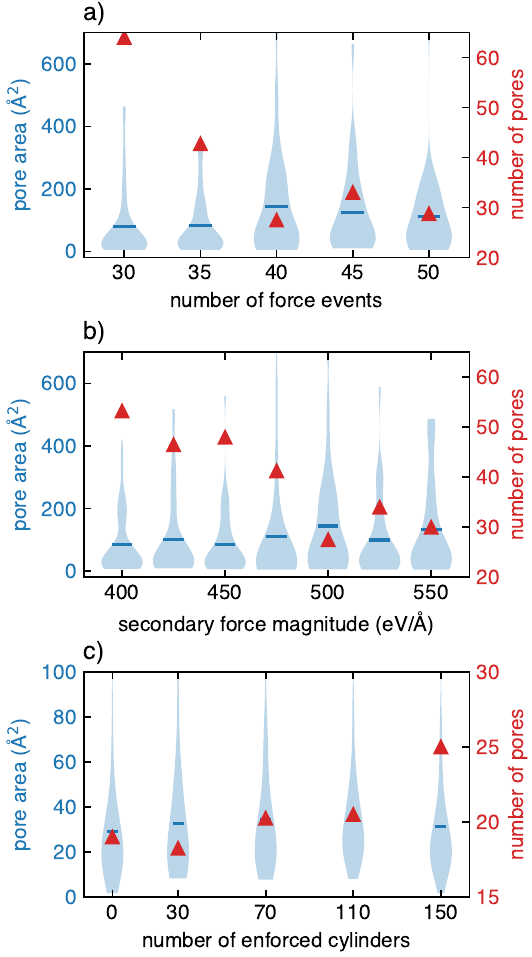}
    \caption{Pore area distributions shown in blue shaded regions (left $y$ axis) for event-varied (a), force-varied (b), and the \SI{4}{ps} annealed region restricted  simulations (panel c), with the total pore count shown in solid red triangles (right $y$ axis). Mean pore area is indicated by horizontal blue lines for each distribution. Note that panels a) and b) have different y-axis scales compared to panel c).}
    \label{fig:pore_stats}
\end{figure}

Figure~\ref{fig:pore_stats} presents the pore analysis data for the event-varied simulations (panel a), force-varied (panel b), and the exclusion cylinder \SI{10}{ps} annealed CNMs (panel c).
Distribution of pore areas shown by the shaded blue violin-shaped
regions.
For context, a pore area of \SI{30}{\angstrom^2} would correspond to a pore radius of about \SI{3}{\angstrom}, assuming a circular pore geometry.
Absolute pore counts are also shown in the same plots marked by the solid red triangles, with values indicated on the right vertical axis.
For both event-varied and force-varied simulations, pore area distributions appear as either skewed normal distributions, or bimodal with a dominant distribution of smaller pores with an area between 5 to \SI{80}{\angstrom^2}, and a minor distribution of larger pores above $\sim\SI{100}{\angstrom^2}$.
The overall range in pore areas are, in general, smaller for the hole-enforced simulations (panel c), which is intuitive given that hole geometry was contrived to an extent.

Considering the event-varied data in Figure~\ref{fig:pore_stats}a), a somewhat counter-intuitive result can be observed. Here, the pore count reduces from roughly 60 and plateaus at 30 for an increasing number of force events, with 40 events seemingly being a threshold for producing fewer but larger pores.
Visual inspection of the trajectories revealed that below the threshold event count of about 40 events, pores were primarily formed due to the overall low carbon atom density in the membrane, resulting in many smaller pores as the carbon structure anneals and develops.
Regions near event sites tended to densify as carbon atoms were pushed away from the locus of the force event.
Hence after a sufficient number of force events the remaining material had enough carbon atoms to feed the structural evolution during annealing, resulting in fewer small pores.
Figure~\ref{fig:pore_stats}b) indicates that pore count also tends to decrease with increasing force magnitude. 
Pore area distributions however, do not seem to follow any obvious patterns, with distributions similar to those seen in Figure~\ref{fig:pore_stats}a).

In the case of the enforced cylinder CNM structures at a simulation time of \SI{9}{ps}, pore area distributions shown in Figure~\ref{fig:pore_stats}c) have a lower spread compared to those in panels a) and b), with pore areas ranging from a few square \AA ngstr\"om, up to about \SI{300}{\angstrom^2}.
For low cylinder count CNMs, the mean pore area is approximately \SI{30}{\angstrom} for all cylinder counts. 
When assuming a circular pore shape, this corresponds to a pore diameter of \SI{5.5}{\angstrom} (measured from carbon nuclei centers, and therefore excluding the effective electron cloud size), similar to the exclusion cylinder radius. 
Although it is interesting for that all exclusion cylinder counts used result in similar final pore counts after relaxation without the exclusion region constraints.

The momentum transfer simulations in general produced broader, and more complex pore area distributions, when compared to the constrained region restricted  CNM structures.
These results suggest that there may be multiple pore creation mechanisms that occur during electron irradiation of the SAMs.
As the conversion process from SAM to CNM is not yet fully understood, it is difficult to determine what form the pore distribution should take.
Unfortunately, information regarding the pore size distributions cannot yet be extracted from HCI charge exchange data.
A deeper understanding of the SAM to CNM conversion process is still required to further understand the pore area distribution within the limitations set by the experimental data.
Future work should explore this aspect of CNM structure, and its influence on the resultant structure-property relationships.
\FloatBarrier

\section{Conclusion}
The atomistic structure of CNMs was explored using a combined experimental and simulation approach.
HCI charge exchange spectra for cleaned CNMs exhibit a characteristic distribution of angle-dependent charge exchange patterns.
Ions exiting the membrane in high charge states are not significantly scattered, whereas ions in low outgoing charge states are deflected more.
The high charge state distribution can be clearly linked to ions passing through thinned and void regions in the membrane, whereas ions in the lower distribution pass through dense material regions~\cite{wilhelm_charge_2014}.
The lower distribution is sensitive to the specific material (areal) density and under the assumption that only a negligible amount of carbon is lost during CNM crosslinking, and hence inhomogeneities in membrane density is an additional independent indicator of sub-nanometer porosity.

Two independent MD simulation techniques were used to produce candidate models of the CNM.
One method explored possible carbon network configurations for differing degrees of porosity, while the other elucidated a possible mechanism of pore formation due to momentum impulses that might occur during precursor SAM irradiation.
We link the candidate atomistic models from MD to the observed charge exchange spectra in HCI transmission with the help of a model based on classical equations of motion of the ions propagating through a given atomic structure.
A quantitative comparison between the experimental and simulated ion transmission spectra with experiment for the different MD structures reveals that a sub-nanometer porous structure with significant fraction of under-coordinated carbon is likely.
This is further corroborated by comparing the predicted tensile moduli to literature data.

The CNM structures that best fit the experimental touch points would likely be chemically reactive at ambient conditions, and could be passivated by hydrogen, as well as water and other oxygen bearing groups, after the CNM has been removed from vacuum. 
This aspect of CNM production is seldom discussed in the literature.
We suggest that structure passivation is a crucial aspect of the CNM, stabilizing the reactive carbon network that is left behind after electron irradiation of the precursor SAM, and may even play a role in the functional properties of the CNM.
Our study is the first to give a quantitative insight into the atomistic structure of CNMs from coinciding experiment and simulation, paving the way to a comprehensive understanding of the mechanical and chemical properties of organic short-range ordered membranes.

\begin{acknowledgement}

We thank the National Computing Infrastructure (NCI) in Canberra, and the Pawsey Supercomputing Centre in Perth, Australia, for provision of computing resource under the NCMAS scheme, as well as the Austrian Scientific Computing (ASC) center for additional computational resources. 
Additionally, we gratefully acknowledge funding from the Austrian Science Fund (FWF) through Grant-DOIs 10.55776/PIN7223324, 10.55776/P36264, 10.55776/I4914, and 10.55776/Y1174.
JS thanks the Deutche Forschungsgemeinschaft (DFG) for financial support through grant number 552519713 (SCHN~615/30).
CN, VS, and AT acknowledge financial support of the DFG (Project TU149/22-1 552519713 and RTG 3014 "PhInt") and BMBF (Project 13N15744 "SARSCoV2Dx")
\end{acknowledgement}

\begin{suppinfo}

Supplementary methods and results are provided as an additional document.

\end{suppinfo}

\bibliography{ref}

\end{document}